\documentclass[runningheads]{llncs}

\usepackage{caption}
\usepackage{graphicx}
\usepackage[table]{xcolor}
\usepackage{tabularx}
\newcolumntype{C}[1]{>{\centering\arraybackslash}p{#1}}

\begin{document}

\title{Measuring Equality and Hierarchical Mobility on Abstract Complex Networks}
\titlerunning{Equality \& Mobility on Networks}

\author{Matthew Russell Barnes\inst{1} \and Vincenzo Nicosia\inst{2} \and Richard G. Clegg\inst{1}}

\authorrunning{M. R. Barnes et al.}

\institute{School of Electronic Engineering and Computer Science, Queen Mary University of London, UK \and School of Mathematical Sciences, Queen Mary University of London, UK}

\maketitle     

\begin{abstract}
The centrality of a node within a network, however it is measured, is a vital proxy for the importance or influence of that node, and the differences in node centrality generate hierarchies and inequalities. If the network is evolving in time, the influence of each node changes in time as well, and the corresponding hierarchies are modified accordingly. However, there is still a lack of systematic study into the ways in which the centrality of a node evolves when a graph changes. In this paper we introduce a taxonomy of metrics of equality and hierarchical mobility in networks that evolve in time. We propose an indicator of equality based on the classical Gini Coefficient from economics, and we quantify the hierarchical mobility of nodes, that is, how and to what extent the centrality of a node and its neighbourhood change over time. These measures are applied to a corpus of thirty time evolving network data sets from different domains. We show that the proposed taxonomy measures can discriminate between networks from different fields. We also investigate correlations between different taxonomy measures, and demonstrate that some of them have consistently strong correlations (or anti-correlations) across the entire corpus. The mobility and equality measures developed here constitute a  useful toolbox for investigating the nature of network evolution, and also for discriminating between different artificial models hypothesised to explain that evolution.

\end{abstract}

\section{Introduction}
\label{sec:introduction}
Hierarchies exist in every facet of our daily lives and understanding their ordering and dynamics is important to discern patterns and enable informed comparisons. A recent paper~\cite{Iniguez2021UniversalRanking} studies how rankings evolve over time and the universal dynamics of hierarchies. In our work we apply a similar approach in the context of networks that evolve in time. The analysis of time evolving networks traditionally consists of tracking the evolution of 
network statistics such as degree distribution, clustering coefficient, assortativity and diameter. However, these overall statistics usually  do not capture how individual nodes change their position within hierarchies, i.e., how their ``status'' improves or deteriorates over time.

In this paper we look at networks evolving in time, from the perspective of individual nodes, and  we study how the characteristics of such nodes change as the network changes, with a special focus on degree centrality.
A node's degree could be considered as a measure of its status or importance
e.g. the number of citations in a citation network~\cite{Redner2005CitationReview} or friends in a social network. Obviously, not all nodes in a network have the same importance, and their roles can change over time as a result of the evolution of the network. Equality is a classical measure to quantify whether nodes tend to have the same centrality as other nodes, or whether centrality is concentrated in some nodes and not others. Similarly, hierarchical mobility measures whether nodes are frozen in their hierarchy position of high or low centrality.
It is particularly interesting to consider how equality and hierarchical mobility evolve as a network changes\footnote{We introduce the term hierarchical mobility to avoid confusion with the word ``mobility'' alone, commonly used in spatial networks as a measure of nodes' ability to move in geographic space.}. 

Consider the Barab\'{a}si--Albert (BA) model~\cite{Barabasi1999EmergenceNetworks} and the Fortunato model~\cite{Fortunato2006Scale-freeRanking} both of which produce power law degree distributions. In the Fortunato model a node attracts new nodes with a probability inversely proportional to the order of its arrival in the network. In BA that probability is proportional to the current degree of the node. Consider an instantiation of a BA model $G_B$ and a separate instantiation of a Fortunato model $G_F$. In both realisations, a node $s$ arriving earlier than a node $t$ has an advantage in gaining links. Imagine in both that $t$ by chance gains more nodes than $s$. In $G_B$ it is likely to continue to gain advantage over $s$ but in $G_F$ the node $s$ will retain its higher likelihood of gaining nodes. In $G_B$, therefore, this change of ranking where $t$ overtakes $s$, if it occurs, is more likely to be permanent. 

Further to this we extend the focus to the importance of an individual node's neighbourhood, and measure this by taking the mean degree of all nodes in said neighbourhood. The interaction between nodes can be somewhat measured by the impact a node has on its neighbourhood's mean degree over time. Consider node $n$ which at $t_1$ is in a neighbourhood with a smaller than average mean degree and at $t_2$ is in a neighbourhood with a larger than average mean degree. Does this growth for the neighbourhood correlate with node $n$ itself having a large degree?

To look at equality we borrow the concept of the Gini coefficient~\cite{Gini1912VariabilitaMutabilita} from classical economics which is generally used to measure wealth or income disparity.
Traditional measures of economic or social mobility proved unsuitable for the setting of time evolving networks,
hence we developed a taxonomy of six hierarchical mobility measures. These measures correlate the degree of individual nodes and their neighbourhoods between different points in time (see section \ref{sec:taxonomy}). 

To consider these measures on real network data we collected a corpus of 30 networks that evolve in time from a number of online sources. The code used to collect this data and analyse it for this paper is freely available under an open source licence\footnote{\label{foo:github}The code and data set collection information is stored here \url{https://github.com/matthewrussellbarnes/mobility\_taxonomy}}.The networks are classified by their field and by structural properties.We used our equality measure and our taxonomy of mobility measures on this corpus to look for common patterns arising in different classes of networks. For example, we correlate the degree of a node at time $t_1$ with its change in neighbourhood mean degree between $[t_1,t_2]$. We call this \emph{philanthropy} as it can be thought of as measuring whether a large degree node helps its neighbours gain degree. Each network in our corpus can be classified by its equality, and the six measures in our mobility taxonomy. We used these seven dimensions to discriminate between different fields of study in complex networks.

We performed a principal component analysis and produced a ``taxonomy landscape'' plot within which types networks can be  differentiated. Further, plotted each taxonomy aspect's change over time for all of the networks in the corpus. Again, this produced plots which show many how the networks can be differentiated based on our taxonomy.

The measures introduced in this paper are a new tool for giving insight into how networks behave. For example, by looking at how network evolve in time over our seven dimension we can see which networks have ossified into a pattern of behaviour and which networks change how they evolve. Most networks in our corpus exhibit a mostly static degree hierarchy but, to our surprise, some exhibited a somewhat mobile degree hierarchy. There were large differences in equality between the networks in our corpus but the majority of networks we studied became more unequal over time. 

\section{Related Work}

\subsection{Individual Influence on Status Hierarchy Evolution}

Abstract networks that evolve in time are an attempt to model the changes in real networks. The beginning of the pursuit for universal dynamics in the time evolution of such growing networks is often attributed to the famous Barab\'{a}si-Albert (BA)~\cite{Barabasi1999EmergenceNetworks} model. Their model replicates power-law degree distributions using simple evolution rules. 

A more abstract approach was taken recently in~\cite{Iniguez2021UniversalRanking} where the universal dynamics of ranking was explored for empirical systems. Through observing the dynamics of the ordered list, the authors built a framework consisting of only two key types; replacement and displacement. Either ranks are ``swapped'' at long ranges in the hierarchy, or ranks ``diffuse'' slowly between close by ranks. Their modelling approach consisted of calculating the probability of every individual to change rank to any other rank, and determined this to be monotonically increasing for ``open'' systems and symmetric for systems which are less open. Openness represents how fast new individuals to the system get into the top 100 ranks.

Approaches focusing on how networks evolve are becoming more numerous and many approaches try to replicate measured evolution dynamics. For instance, both~\cite{Fire2020TheTime} and~\cite{Nsour2020Hot-Get-RicherModel} take the approach of measuring the effect node longevity has on its degree. It was found that a model which mimicked the preferential attachment of BA, but only for newly arrived nodes, did well in replicating the observed relationship between nodes and edges. 

Another example is~\cite{Zhou2020UniversalNetworks} which focused on mimicking the evolution of assortativity in social networks by calculating the probabilities of every node to be connected to $k$ neighbours at time $t$. From this they derived an analytical model for calculating the assortativity of a network throughout its lifetime, and found the model was a good fit with real social network data sets.

\subsection{Equality}
Determining the distribution of finite resources among many actors in a system is a useful metric for understanding whether such resources are being concentrated or spread out. In both sociology and economics this has been a popular area of interest with each concentrating on the distribution of ``status'' in societies. Economists use financial income as a quantifiable proxy for status, whereas sociologists tend to keep the definition more broad and qualitative.

A common measure of equality used by economists is the Gini coefficient~\cite{Gini1912VariabilitaMutabilita} which is derived with reference to the Lorenz curve~\cite{Lorenz1905MethodsWealth}, a plot of wealth versus population. The Gini coefficient $G$ is given by
\begin{equation}
\label{eq:gini}
    G = \frac{\sum^{n}_{i=1}\sum^{n}_{j=1}|x_{i}-x_{j}|}{2n^2\overline{x}}
\end{equation}
where $x_{i}$ is the income of person $i$, $n$ is the number of people, and $\overline{x}$ is their mean income. This makes tracking changes in the equality measurements of populations a simple calculation. However, it does remove nuance from the result as it is a single number.

In sociology, one growing area of research has been network effects on inequality of status in social networks~\cite{Granovetter1973TheTies}. Much of the focus is aimed at how peer influence and network homophily~\cite{Dimaggio2012NetworkInequality} are large contributors to contagion of ``practices'' which improve a person's status.

Our approach is to measure the Gini coefficient of the distribution of degree as it evolves over time.

\subsection{Hierarchical Mobility}
Tracking the mobility of status hierarchies over generations has also been a focus of Sociologists and Economists. Hierarchical mobility is the amount of movement individuals experience between hierarchy levels, and the hierarchies chosen depend on the focus of interest. 

Sociological hierarchies focus on occupational classes~\cite{Szreter1984TheOccupations} which are qualitative in nature and therefore have only a subjective~\cite{Erikson1992TheFlux} direction of hierarchy so do not apply to our purely network topology approach. An example mobility measurement is the ``Log-Multiplicative Layer Effect Model''~\cite{Breen2004SocialEurope} which compares two matrices (called ``layers'' or ``generations'') of class associations by assuming a uniform multiplicative association. This uniform association removes much of the much needed nuance between inter-generational class associations. 

Economic hierarchies are built out of financial income bands~\cite{Mayer2005HasChanged} which are quantifiable and so readily ranked. A widely used mobility measurement used in economics is the Pearson correlation coefficient~\cite{Solon2001IntergenerationalMobility}
$\beta = r(\ln{Y_c},\ln{Y_p})$
where $r(X,Y)$ is the Pearson correlation between numerical series $X$ and $Y$. This is usually used in conjunction with
$\ln{Y_c} = \alpha + \beta_p\ln{Y_p} + \epsilon_c$
where $Y_c$ is the income of children, and $Y_p$ is the income of parents, $\alpha$ is a constant and $\epsilon_c$ is a fitted constant.

Our proxy for status in abstract networks is degree centrality which, like income bands, has a definite hierarchical direction. Instead of imposing arbitrary generations we study the evolution between two points in time $t_1$ and $t_2$. 
In section \ref{sec:taxonomy}, we introduce more details of translating these hierarchical mobility measurements into our taxonomy of hierarchical mobility.

\section{Measuring equality and hierarchical mobility in networks}
\label{sec:taxonomy}

\subsection{Equality}
Using equation \ref{eq:gini} we can substitute in the degree of each node as $x_i$, the mean degree as $\overline{x}$ and the number of nodes as $n$. This is done at many time-steps throughout the life of a network and then the resulting coefficients are plotted against time to see how equality changes. 

The income hierarchies used in economic analysis are very similar to degree in that they are numeric and a larger number is assumed to be a good proxy for status. However, the income of individuals fluctuate over their lifetime, both gaining and losing income. Our  formulation of growing networks contain nodes that only ever increase their degree. This does not affect the calculation of equality, and moreover it allows for greater inference.

If a network is gaining a larger equality over time then the range of degree of nodes is getting narrower. As the networks are always growing it can be inferred in this case that high degree nodes are connecting to low degree more preferentially. Conversely, a lowering of equality over time suggests a divide in the connections being created. It can be inferred that lower degree nodes are more likely to attach to higher degree nodes, and less likely between themselves.

\subsection{Mobility Taxonomy}
As mentioned earlier, classical notions of social and economic mobility were not a good fit for the context of time evolving networks. To this end we introduced new measures that depend on the centrality of a node and how it evolves. In this work we consider only degree centrality. 

To calculate the mobility of a node we consider the degree centrality of every node at time $t_1$ and the change of centrality in the period $[t_1,t_2]$. If these are highly correlated it indicates that the nodes with highest centrality at $t_1$ gain the most in the period to $t_2$. We measure this correlation using the Pearson correlation coefficient and refer to this measure as mobility\footnote{Technically this should be called ``anti''-mobility as a larger correlation coefficient refers to fewer changes in hierarchical position.}.

We also extended this to look at the mean centrality of each node's neighbours at time $t_1$ and the mean gain of centrality for this same neighbour set in the period $[t_1,t_2]$. This gives us four measures: node degree, change in node degree, mean neighbourhood degree and change in mean neighbourhood degree. Looking at the correlations between each of these gives us a taxonomy of six different mobility measures which are tabulated below.

\vspace{1em}
\begin{minipage}{\textwidth}
\begin{tabular}{| C{2.7cm} | C{2.7cm} | C{2.7cm} | C{2.7cm}|} \hline
\cellcolor{yellow} Correlation & \cellcolor{yellow} Change in degree & \cellcolor{yellow} Mean neighbour degree & \cellcolor{yellow} Change in mean neighbour degree \\ \hline
\cellcolor{yellow} Degree \vskip 0cm & Mobility & Assortativity & Philanthropy \\ \hline
\cellcolor{yellow} Change in degree & \cellcolor{lightgray} & Community & Change in assortativity \\ \hline
\cellcolor{yellow} Mean neighbour degree & \cellcolor{lightgray} &  \cellcolor{lightgray} & Neighbour mobility \\ \hline
\end{tabular}
\captionof{table}{The taxonomy of mobility related aspects}
\end{minipage}
\vspace{1em}

Correlating degree at time $t_1$ with the change in average neighbourhood degree to $t_2$ can be thought of as measuring how much a high status node helps its neighbours and we call this \emph{philanthropy}. The correlation between neighbour mean degree at $t_1$ and the change in the period $[t_1,t_2]$ can be thought of as \emph{neighbourhood mobility}. The correlation between a neighbourhood's mean degree at time $t_1$ and a node's gain in degree between $[t_1,t_2]$ could be considered as a measure of how much neighbourhood helps a node and we refer to it as \emph{community}. The correlation between a node's degree and its neighbours' mean degree at $t_1$ is the well known \emph{assortativity}. The correlation between a node's change of degree and its neighbours' mean degree between $[t_1,t_2]$ can therefore be thought of as \emph{change in assortativity}\footnote{This is not quite correct as the assortativity at $t_2$ would be measured on the neighbourhood set at $t_2$ not the neighbourhood set at $t_1$.}. These six measures together form a taxonomy for investigating how individual nodes and their neighbours interact and change over time.

\section{Data set Corpus}
To empirically measure the prevalence of each aspect of our taxonomy in many different types of network, we have collected 30 real networks. We have assigned each data set a type based on where the data has been collected, i.e social or transportation, and information about where to find them is available under an open source licence \footnote{See footnote \ref{foo:github}}. Furthermore, we assigned each a category taken from the structural characteristics of the networks themselves, such as bipartite or the nature of newly joining nodes. For instance, citation networks grow by adding stars every iteration, i.e one node connected to all those it cites. The networks vary in size from 138 edges to nearly 1 million~\footnote{Some networks were originally above 1 million edges but here are truncated to keep computational time reasonable}. All networks are treated as undirected and unweighted with added links never removed.

\begin{table}[!tbp]
    \resizebox{\textwidth}{!}{\begin{tabular}{c|c|c|c|c|c|c}
        & Name & Description & Type & Structure & Nodes & Edges  \\
         \hline \hline
        a & College Messages \cite{Panzarasa2009PatternsCommunity} & Messages between students on a UC-Irvine message board & Social & Star & 1,881 & 13,695 \\
        \hline
        b & SCOTUS Majority \cite{Fowler2007NetworkCourt} & Legal citations among majority opinions by SCOTUS & Citation & Star & 33,442 & 199,374 \\
        \hline
        c & Amazon Ratings \cite{Lim2010DetectingBehaviors} & Amazon user connects to all products they have rated & Economic & Bipartite & 743,018 & 975,429 \\
        \hline
        d & Apostles Bible \cite{Holanda2019CharacterClassification} & Characters in the Holy Bible’s Acts of Apostles & Co-occurrence & Individual & 75 & 160 \\
        \hline
        e & Appollonius \cite{Holanda2019CharacterClassification} & Characters in The Life of Apllonius of Tyana & Co-occurrence & Individual & 92 & 138 \\
        \hline
        f & Citations US Patents \cite{Hall2001TheTools} & Citations among patents in the United States & Citation & Star & 777,527 & 999,632 \\
        \hline
        g & Classical Piano \cite{Park2020NoveltyNetworks} & Transitions of chords in western classical piano music & Co-occurrence & Individual & 141,571 & 501,033 \\
        \hline
        h & Email EU \cite{Paranjape2017MotifsNetworks} & E-mails between users at an EU research institution & Social & Star & 986 & 16,006 \\
        \hline
        i & Procurement EU \cite{Wachs2021CorruptionPerspective} & Public EU procurement contracts & Economic & Bipartite & 330,049 & 566,369 \\
        \hline
        j & Facebook Wall \cite{Viswanath2009OnFacebook} & Posts by users on other users' Facebook wall & Social & Star & 45,580 & 181,666 \\
        \hline
        k & Lord Of the Rings \cite{Clegg2009} & Character co-occurrence in Lord of the Rings Trilogy & Co-occurrence & Individual & 139 & 634 \\
        \hline
        l & Luke Bible \cite{Holanda2019CharacterClassification} & Characters in the Holy Bible’s Luke Gospel & Co-occurrence & Individual & 76 & 203 \\
        \hline
        n & PhD Exchange \cite{Taylor2017Eigenvector-basedNetworks} & Exchange of PhD mathematicians between unis in the US & Citation & Star & 230 & 3,643 \\
        \hline
        o & Programming Languages \cite{Valverde2015PunctuatedLanguages} & Influence relationships among programming languages & Citation & Star & 366 & 759 \\
        \hline
        p & Reuters Terror News \cite{Corman2006StudyingSystems} & Word co-use in Reuters 9/11 coverage & Co-occurrence & Individual & 13,265 & 146,985 \\
        \hline
        q & Route Views \cite{Clegg2009} & Route Views internet topology & Computer & Individual & 33,644 & 94,075 \\
        \hline
        r & Reddit Hyperlinks Body \cite{Kumar2018CommunityWeb} & Subreddit-to-subreddit hyperlinks from body of posts & Social & Clique & 35,592 & 123,394 \\
        \hline
        s & Reddit Hyperlinks Title \cite{Kumar2018CommunityWeb} & Subreddit-to-subreddit hyperlinks from title of posts & Social & Individual & 53,747 & 217,986 \\
        \hline
        t & Hospital \cite{Vanhems2013EstimatingSensors.} & Contacts between everyone in a hospital ward & Contact & Spatial & 75 & 1,132 \\
        \hline
        u & Hypertext Conference \cite{Isella2011WhatsNetworks} & Contacts among attendees of ACM Hypertext 2009 & Contact & Spatial & 113 & 2,192 \\
        \hline
        v & Infectious \cite{Isella2011WhatsNetworks} & Contacts during Infectious SocioPatterns 2011 event  & Contact & Spatial & 10,844 & 43,951 \\
        \hline
        w & Office \cite{Genois2015DataLinkers} & Contacts between individuals in an office building & Contact & Spatial & 92 & 754 \\
        \hline
        x & Primary School \cite{Stehle2011High-resolutionSchool} & Contacts among students and teachers at a primary school & Contact & Spatial & 242 & 8,298 \\
        \hline
        y & AskUbuntu \cite{Paranjape2017MotifsNetworks} & User answers or comments on questions on AskUbuntu & Social & Clique & 157,709 & 502,966 \\
        \hline
        z & MathOverflow \cite{Paranjape2017MotifsNetworks} & User answers or comments on questions on MathOverflow  & Social & Clique & 24,506 & 198,040 \\
        \hline
        A & StackOverflow \cite{Paranjape2017MotifsNetworks} & User answers or comments on questions on StackOverflow  & Social & Clique & 38,379 & 618,519 \\
        \hline
        B & SuperUser \cite{Paranjape2017MotifsNetworks} & User answers or comments on questions on SuperUser  & Social & Clique & 124,528 & 541,466 \\
        \hline
        C & UCLA AS \cite{Clegg2009} & UCLA AS level internet topology & Computer & Individual & 38,055 & 224,545 \\
        \hline
        D & US Air Traffic \cite{Paranjape2017MotifsNetworks} & Flights among all commercial airports in the US & Transport & Individual & 623 & 14,952 \\
        \hline
        E & Wiki Talk \cite{Paranjape2017MotifsNetworks} & Wikipedia users editing each other's Talk page  & Social & Individual & 116,661 & 329,805 \\
    \end{tabular}}
        \caption{Data sets used for network creation, limited to 1 million edges.}
    \label{tab:datasets}
\end{table}

\section{Results \& Discussion}

\subsection{Equality}
\begin{figure}[!tbp]
\centering{
  \includegraphics[width=0.8\textwidth]{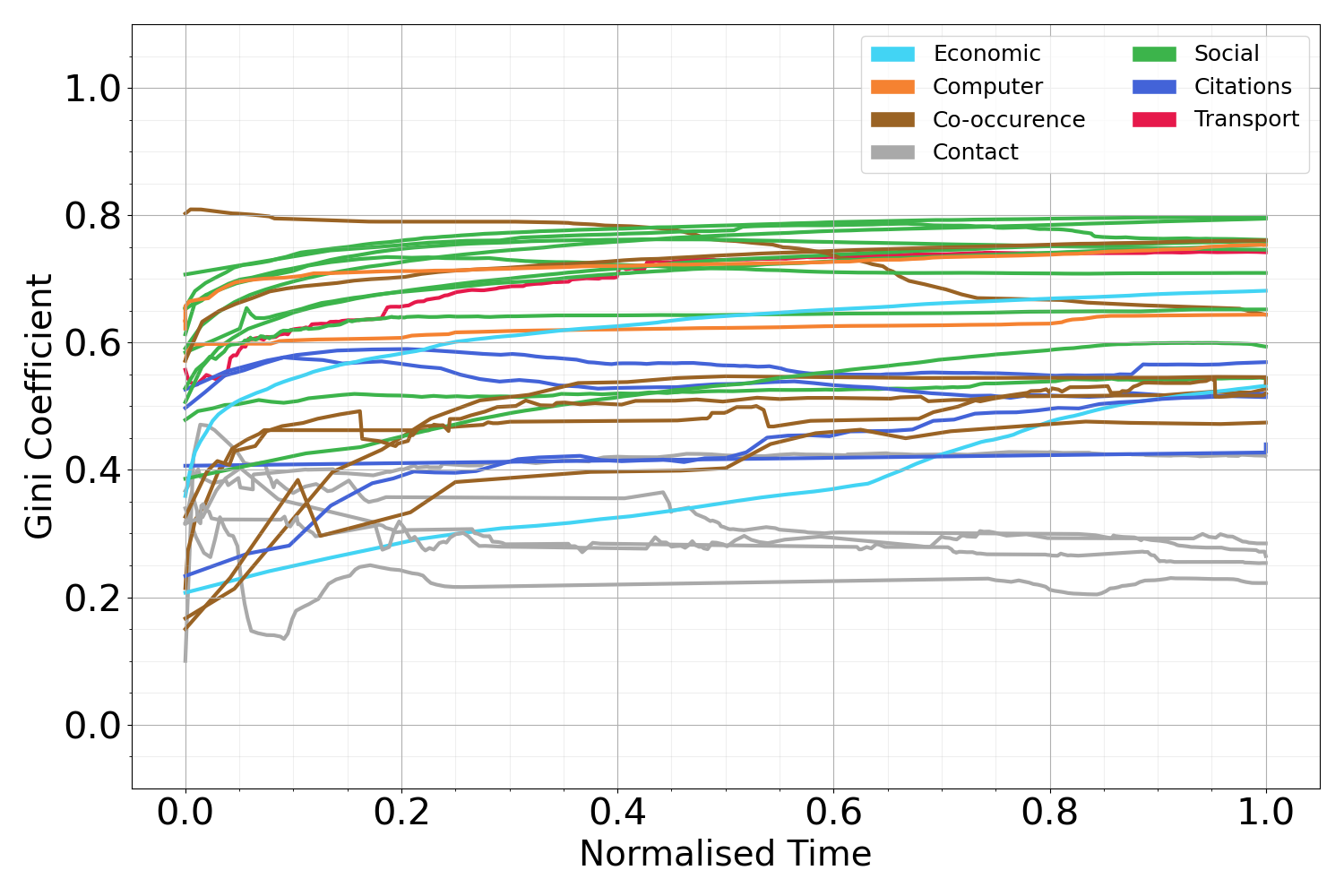}
  \caption{Equality of data set corpus networks coloured by data type over normalised time, where all network lifetimes are normalised to the range $0$ to $1$.}
  \label{fig:equality_time}
  }
 \end{figure}

We plot the Gini Coefficient (GC) of all our corpus networks on the same axis and colour the data set based on the datatype shown in Table~\ref{tab:datasets}. The x-axis in figure~\ref{fig:equality_time} is normalised time where the maximum duration of each network has been normalised to be between [0,1], and 100 snapshots of equality have been calculated throughout the network's lifetime. The y-axis shows the GC, where a higher value corresponds to a lower level of equality for the network.

In Figure 1 the GC takes values over most of the possible range (0,1) with the values being contained between 0.2 and 0.8 from normalised time 0.2 onwards. Of the 30 data sets studied all but 4 rise in GC (becoming more unequal) between normalised time 0.0 and 1.0. he figure also shows some discrimination between the types of networks, for example social networks in general are most the unequal and contact networks most equal.

The contact networks are collected by measuring contact between people in physical space. Therefore the degree of a node is limited by the number of people that can be physically present in the space during the measurement interval. These limitations on network evolution would suggest a more equal distribution of degree as it is more likely that two individuals will spend time together than a less constrained system. Online social networks do not have this physical limitation. For example, if an individual is not interested in the software language Python, then they will not interact with the Python section of stackoverflow.com. This means there is less chance of densification as the interactions are less likely to occur than a physical contact network.

\subsection{Mobility Taxonomy}

To look at how the mobility taxonomy varies with time for every data set network we chose values of $t_1$ evenly spaced as the network grows. To achieve this we pick values of $t_1$ corresponding to $10, 20, \ldots, 90 \%$ of the final number of edges being present and $t_2$ corresponding to all edges being present.

\subsubsection{Taxonomy Aspect Evolution}
\begin{figure}[!tbp]
\centering{\includegraphics[width=1.0\textwidth]{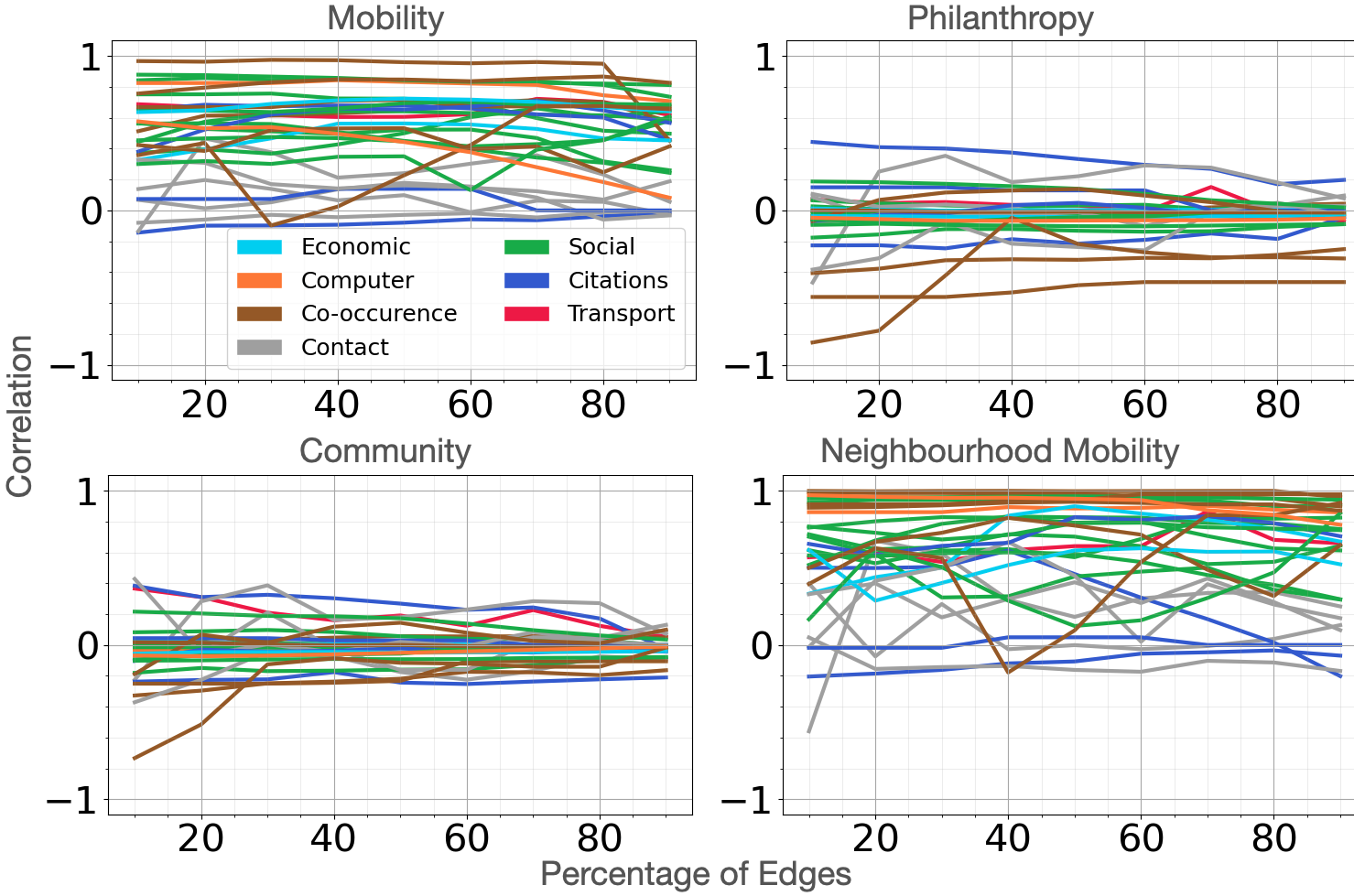}
  \caption{Every data set corpus network for each aspect of the mobility taxonomy coloured by data type. The x-axis represents $t_1$ characterised as percentage of life of the network. The y-axis shows the correlation coefficient as measured using Pearson correlation.}
\label{fig:taxonomy_over_time}
}
\end{figure}

These values of the mobility taxonomy were plotted against time to show how they individually evolve with time in figure \ref{fig:taxonomy_over_time}. The data set types are again used to differentiate the networks, and it can be seen that most of the networks have a similar evolution for each of the aspects. However there are notable exceptions in each case.

Mobility and neighbourhood mobility positively correlate with a coefficient of 0.79 at $10\%$ of edges, which can be seen in the vast majority of networks reaching high levels of correlation for both aspects. Positive correlation coefficients for these aspects signify a more static hierarchy of degree in the network over time (i.e individual nodes rarely change their place in the degree hierarchy). As static degree hierarchies are sometimes thought of as a given in real world network degree, it is unexpected we found limited evidence of networks which have a somewhat mobile degree  hierarchy. However, this is largely in contact networks where, perhaps the large degree nodes happened to have many contacts at one time and getting the mobile degree hierarchy is merely regression to the mean.

The aspects community and philanthropy have a Pearson correlation of 0.48 at $10\%$ of edges. Some networks are positive or negative in both but most of the networks show very little either way. For those on the periphery, there is a noticeable trend towards zero for both positive and negative correlations suggesting a longer time in the network correlates with a more significant influence from these aspects. Negative values of philanthropy can be thought of as nodes that grow themselves but have a detrimental effect on the growth of their neighbours. This is largely seen in co-occurrence networks and we might think of a ``prima-donna'' effect where characters that gain attention drain attention from their co-stars.

\subsubsection{Principal Component Analysis}
To take the analysis further we correlated each aspect with each other, plus the Gini coefficient results at time $t_1$ for completeness, e.g all of the data set results for mobility with those of philanthropy. This resulted in a seven-dimensional matrix which we reduce to its two highest variance components using Principal Component Analysis for visualisation

\begin{figure}[!tbp]
\centering{
  \includegraphics[width=0.7\textwidth]{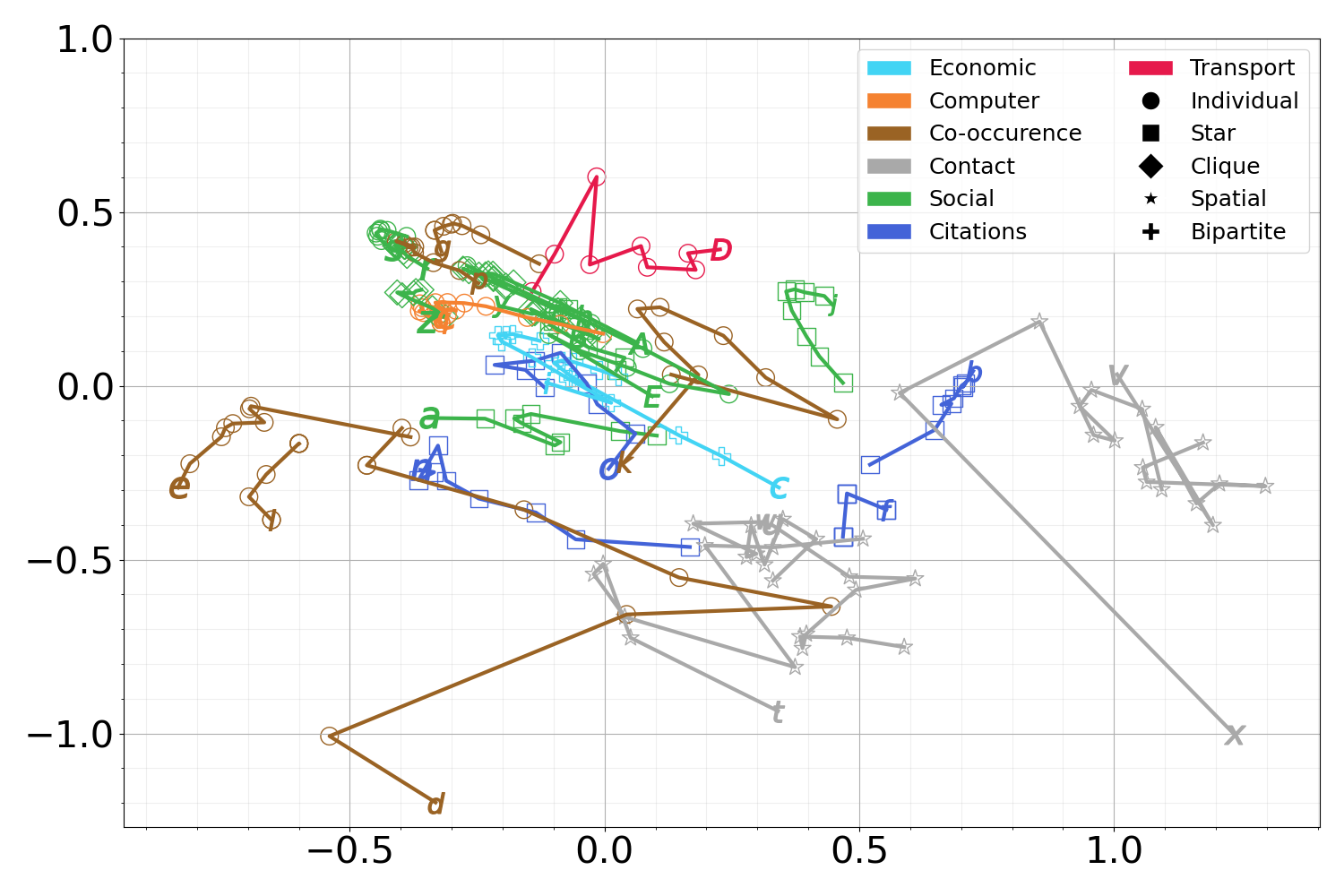}
  \caption{Principal component analysis of the mobility taxonomy, including equality, with each data set marked by its associated letter (see Table \ref{tab:datasets}) on its first time-step. The plotted lines denote the data sets through 9 time-steps, each step plotted with a shape corresponding to its structure and the colour corresponding to its type.}
  \label{fig:PCA}
}
\end{figure}

Having time dependant values for the PCA means we are able to show how the networks evolve in time. These results are plotted in figure \ref{fig:PCA} with the data and structural types from Table \ref{tab:datasets} highlighted using colours and shapes. 
One striking feature is that many network are clustered together in the upper left quadrant of the graph.
These networks are also much less prone to large changes in position, i.e they are more ossified, than those on the periphery of the cluster. A slight trend occurs in how the further the network is from the origin the more volatile its position on the PCA is over time. This volatility of position also correlates with a fewer number of nodes in the networks, i.e smaller networks have more changeability in their taxonomy aspects over time.

\section{Conclusions}
\label{sec:conclusion}

In this paper we have shown how tracking the statistics of individual nodes and their neighbours, specifically the degree centrality, throughout the lifetime of the network can bring structural and developmental insight into the network's evolution. Each of the mobility taxonomy aspects, along with equality, show different mechanisms underlying the evolution of a network and combining knowledge gained from each of them draws a vivid picture of how individual nodes interact with each other.

Also, running a PCA on the whole mobility taxonomy (plus equality) was shown to distinguish between networks into somewhat distinct categories. The time evolving PCA allowed for determination of network ossification and it was found that social networks are more likely to be ossified whereas co-occurrence and contact networks are less likely, though network size is also a contributing factor.

The techniques outlined in this paper are of great use for analysing trends of nodes in networks which evolve with time. Our main focus for future expansion is to delve deeply into the inter-node interaction dynamics of a single network and build understanding of causation for network-level phenomena from individual node interactions. 

\subsubsection{Acknowledgements}
The authors wish to acknowledge the support of Moogsoft Ltd. for funding this research.

\bibliographystyle{splncs04}
\bibliography{complenet_2022.bib}

\end{document}